\documentclass[cits]{PoS}

\usepackage{amsmath}

\title{Light hadron spectroscopy and pseudoscalar decay constants}

\ShortTitle{Hadron spectrum}

\author{\speaker{Christian Hoelbling}\\
        Bergische Universit\"at Wuppertal, Gaussstr.\,20, D-42119 Wuppertal, Germany\\
        E-mail: \email{hch@physik.uni-wuppertal.de}}

\abstract{I review recent lattice QCD results on light hadron
spectroscopy and pseudoscalar decay constants.}

\FullConference{The XXVIII International Symposium on Lattice Field Theory, Lattice2010\\
		June 14-19, 2010\\
		Villasimius, Italy}

\begin{document}

Since nearly four decades QCD is believed to be the fundamental theory
of the strong interaction. Its fundamental degrees of freedom are
quarks and gluons, none of which are experimentally observed. Instead,
one observes a large number of bound state hadrons that are believed
to be the low energy degrees of freedom of QCD. These objects are
inherently nonperturbative. Computing their properties - especially
their masses - therefore constitutes one of the most fundamental
challenges for lattice QCD as the prime method of treating QCD in the
nonperturbative regime.

During the last couple of years there has been enormous progress towards
reaching this goal. Driven by conceptual and algorithmic advances as
well as advances in computer technology, we are currently able to
compute ground state light hadron masses to within a total accuracy of a
few percent in $2+1$ flavor QCD. Our understanding of excited states is
not as detailed as for the ground states, but huge qualitative progress
has been made.

In the current review, I will summarize the state of the art in light
hadron (i.e. hadrons composed of $u$, $d$ and $s$ valence quarks)
spectroscopy. I start with a discussion of systematic uncertainties
(sect.~\ref{sys}) followed by a brief overview of the currently
available data sets (sect.~\ref{ens}). I then review ground state
spectroscopy (sect.~\ref{gs}) including recent attempts to treat
electromagnetic effects (sect.~\ref{em}) and excited state spectroscopy
(sect.~\ref{es}). I close with a section on the closely related light
pseudoscalar decay constants (sect.~\ref{fp}).

A very similar mix of topics has been reviewed at the lattice 2009
conference \cite{Scholz:2009yz}. A more detailed recent
review about excited state spectroscopy can be found
in~\cite{Peardon:2010zz}. A review of recent ensembles generated by
various collaborations is given in~\cite{Jung:2010jt}.

\section{Systematic errors}
\label{sys}

Every lattice calculation of the light hadron spectrum generically
contains a number of extrapolations. Typically the three most important
ones are continuum, infinite volume and chiral extrapolations.  While
the continuum extrapolation is unavoidable, the chiral extrapolation can
be circumvented by directly going to or reweighting to the physical
point. Finite volume effects for stable particles are exponentially
suppressed in the lattice size and can therefore in principle be made
very small. 

In addition current lattice calculations of the light hadron spectrum
typically contain a number of controlled approximations that can
systematically be corrected for by non-lattice methods. For the ground
state hadron spectrum and the pseudoscalar decay constants the two most
important ones are the isospin limit and the treatment of QED
effects. The size of these effects is given by the $\pi^0-\pi^\pm$ mass
splitting. Although it is less common today, one may additionally have
uncontrolled approximations such as the quenched approximation.

Furthermore, depending on the action used and the specific observable,
one may have to account for additional systematic effects such as
flavor/isospin breaking or level shifts of resonant states. I will
briefly summarize here, how the three main sources of systematic errors
affect todays light hadron spectrum and decay constant calculations.

\subsection{Continuum extrapolation}

A final continuum extrapolation is an unavoidable part of any lattice
calculation that wants to make a statement about the underlying
fundamental continuum theory. The severity of the continuum
extrapolation however depends very strongly on both the action used and
the combination of scale setting observable and measured observable.

While gluonic actions used today are generically a variant of $O(a^2)$
Symanzik-improved actions and show a continuum scaling of at least
$O(\alpha_s a^2)$, the scaling behavior of the various fermion actions is
typically not as good. Formally, staggered fermions and twisted mass
fermions at maximal twist as well as exactly chiral fermions show
$O(a^2)$ continuum scaling while Wilson type fermions generically start
with $O(a)$ scaling. Only improved staggered fermions such as asqtad have a
leading scaling behavior of $O(\alpha_s a^2)$. There are several
caveats to this statement however.

For twisted mass fermions, $O(a^2)$ scaling is only strictly realized
at maximal renormalized twist \cite{Frezzotti:2003ni,Aoki:2004ta},
which requires the tuning of one additional parameter. This tuning is
routinely done as part of any twisted mass calculation (see
e.g. \cite{Baron:2011sf}). Since this tuning has a typical accuracy on
the few percent level, it is expected that the $O(a^2)$ terms - though
formally subleading - are in fact numerically dominant. In addition,
twisted mass calculations often employ a doublet of valence fermions
with opposite Wilson parameter to cancel remnant $O(a)$ effects.

Similarly $O(a^2)$ scaling is only strictly realized for chirally
symmetric fermions if the chiral symmetry is exact. Fermion
formulations that incorporate an inexact chiral symmetry, such as
domain wall fermions, do formally have a remaining $O(\alpha_s^n a)$
scaling behavior. The smallness of the residual mass and other
numerical evidence \cite{Aoki:2010dy} however suggest that, similar to
the twisted mass case, the $O(a^2)$ term is dominant although it is
formally subleading.

Wilson type fermions on the other hand are typically Symanzik improved
by the addition of a Sheikoleslami-Wohlert (clover) term. At tree level
($c_{SW}=1$), this results in an $O(\alpha_s a)$ scaling of on-shell
observables while with a suitable nonperturbative tuning one can in
principle obtain $O(a^2)$ scaling. In addition, there is numerical
evidence \cite{Hoffmann:2007nm,Durr:2008rw,Kurth:2010yk,Durr:2010aw}
that the scaling behavior of clover fermions is substantially
improved by gauge link smearing, which is commonly used today.

Apart from the action used, the continuum scaling is also largely
dependent on the observables considered. In lattice QCD all
observables (as well as all input parameters) are dimensionless
quantities. Therefore, if one is interested in dimensionful quantities
such as the baryon masses, the scaling might be quite different
depending on the scale setting variable used. Ratios of baryon masses
generally have shown a very mild scaling behavior and consequently
setting the scale by one baryon mass
\cite{Durr:2008zz,Alexandrou:2009qu} or a combination of baryon masses
\cite{Bietenholz:2010jr,Bietenholz:2010si} is a typical procedure in
calculating light baryon spectra and related quantities
\cite{Capitani:2009tg,Brandt:2010ed}. One can even go further and use
dimensionless mass ratios to a scale setting mass throughout the
analysis \cite{Durr:2008zz}. A slightly different approach is the
scale setting by the MILC collaboration which uses $r_1$
\cite{Bazavov:2009bb}.

Some care has to be taken about the size of the scaling window. While
generally scaling is not expected to set in for lattice spacings coarser
than $a\sim 0.1-0.15\text{~fm}$, it has been observed
\cite{DelDebbio:2002xa,Schaefer:2010hu,Antonio:2008zz,Bazavov:2010xr}
that for fine lattices the autocorrelation time of the topological
charge is rapidly growing. It therefore seems to be prohibitively
expensive with current algorithms to obtain a sufficiently large and
statistically independent ensemble of configurations for lattice
spacings finer than $a\sim 0.05\text{~fm}$.

Generally, for the observables considered in this review continuum
scaling is rather mild and not a leading source of systematic error.

\subsection{Reaching the physical point}

Historically, extrapolating lattice results to physical pion masses
has been one of the leading sources of systematic error in lattice
calculations. While it is often referred to as chiral extrapolation,
the term does not accurately reflect the current state of the art
anymore. Physical pion masses have been reached both by reweighting
techniques \cite{Aoki:2009ix} and by direct simulation
\cite{Durr:2010aw}\footnote{With staggered fermions, the physical
  point has also been reached in the sense that the lowest
  pseudoscalar mass is at the physical pion mass
  \cite{Aoki:2005vt,Aoki:2006br,Aoki:2006we,Aoki:2009sc,Borsanyi:2010bp,Borsanyi:2010cj}}. Other
results have been obtained with pion masses close to the physical one
such that the remaining extrapolation is tiny and does not
substantially contribute to the systematic error anymore
\cite{Durr:2008zz,Collins:2011gw,Bazavov:2010hj}.

In general, there are two strategies of obtaining results at the
physical point: On the one hand, one can extrapolate or interpolate to
the physical point using either a chiral expansion around the singular
point $m=0$ or a regular Taylor expansion around a finite
mass. Alternatively, one may tune the parameters or reweight existing
ensembles to reach the physical point exactly. The latter strategy has
been followed by the PACS-CS collaboration \cite{Aoki:2009ix} and, to a
lesser extent, by the RBC-UKQCD collaboration \cite{Aoki:2010dy} who
reweighted their ensembles to the physical strange quark mass,
performing a subsequent extrapolation to the physical pion mass.  All
other results discussed here use a form of extrapolation or
interpolation to obtain results at the physical point.

For pseudoscalar decay constants, chiral perturbation theory ($\chi$PT)
\cite{Weinberg:1978kz,Gasser:1983yg,Gasser:1984gg} presents a natural
framework to describe lattice data. It is an effective field theory
description based on the chiral symmetry properties of QCD and treats
pseudoscalar mesons as fundamental fields. It represents an asymptotic
series expansion around either the SU(2) or SU(3) chiral point. A
variant where the kaon is treated as a heavy rather than a chiral
particle is known as heavy kaon $\chi$PT
\cite{Allton:2008pn}. Refinements of $\chi$PT take discretization terms
of specific lattice actions
\cite{Bernard:2001yj,Bar:2003mh,Bar:2005tu} and (partial)
quenching
\cite{Sharpe:1992ft,Bernard:1992mk,Bernard:1993sv,Sharpe:2001fh} into
account explicitly.

As an effective theory, each order of the chiral expansion carries with
it a number of expansion parameters (low energy constants or
LECs). Explicit expressions for meson masses and pseudoscalar decay
constants up to NNLO are available in the continuum
\cite{Bijnens:2004hk,Bijnens:2005ae,Bijnens:2005pa,Bijnens:2006jv} while
for lattice formulations only NLO expressions are available.

In order to fix the LECs different strategies are employed. One may
either use the full information available on the required LECs to
constrain the fit, use partial information or let the fit determine the
LECs and then optionally check for the consistency of the result with
other determinations. In fits where LECs are fully constrained the
chiral description of the data is sometimes not satisfactory. In such
cases ``analytic'' higher order terms are often used. These are ad-hoc
polynomial higher order terms introduced for the sole purpose of
obtaining a reasonable fit. These functional forms may be viewed as a
hybrid between a chiral and a regular Taylor expansion that will be
discussed below.

For masses other than the pseudoscalars, $\chi$PT generically predicts a
leading nonanalytic term of the form $M_{\text{PS}}^3$
\cite{Langacker:1974bc}. More formally, one can include baryons in
$\chi$PT \cite{Gasser:1987rb} but the resulting series is only slowly
converging. An alternative formulation with better convergence
properties is heavy baryon $\chi$PT \cite{Jenkins:1990jv,Bernard:1992qa}
which treats baryons as nonrelativistic particles and currently is most
commonly used to fit lattice baryon data. Recently, the covariant
approach \cite{Becher:1999he} that promises better convergence behavior
for heavier pion masses has been revived
\cite{Dorati:2007bk,Durr:2010ni} and used for chiral fits of the baryon
octet.

An alternative to the chiral expansion for describing the pion mass
dependence of any hadronic observable is a Taylor expansion around a
finite pion mass. In contrast to the chiral expansion, the Taylor
expansion is performed around a nonsingular point and consequently has
a finite radius of convergence. Typically this convergence radius is
given by the distance of the expansion point to the chiral
limit. Usually, the expansion is performed in powers of the
pseudoscalar mass square $M_{\text{PS}}^2-M_{\text{PS}_0}^2$, but the
chiral behavior of baryon masses has also been successfully described
by a linear expansion in $M_{\text{PS}}$
\cite{WalkerLoud:2008bp}. Optimal convergence is achieved in principle
by placing the expansion point at the middle of the intervall spanned
by all simulation points and the physical point (for further
discussion of this point see \cite{Durr:2008zz,Lellouch:2009fg}).
Note that from a practical perspective the choice of the expansion
point $M_{\text{PS}_0}^2$ does not play a role in the fit itself as a
redefinition of $M_{\text{PS}_0}^2$ may be absorbed by redefining the
lower order fit coefficients.

It is also worth noting that one may try to fit ratios
\cite{Durr:2008zz} or differences
\cite{Bietenholz:2010jr,Bietenholz:2010si} of baryon masses in order to
cancel common contributions and obtain a more regular chiral behavior.

Recently, SU(3) breaking effects in baryon multiplets have also been
studied in the $1/N_c$ expansion \cite{Jenkins:2009wv} offering an
alternative way of describing the chiral behavior of baryon mass
multiplets.

\subsection{Finite volume effects}

The third common source of error in a lattice calculation is the
finiteness of the volume of the simulated box. As is the case for the
continuum limit, the infinite volume limit can never be reached and an
extrapolation in the volume is in principle unavoidable. For most
observables however the leading finite volume corrections are
exponentially small in the box size and not polynomially and can
therefore be made arbitrarily small by increasing the volume
\cite{Luscher:1985dn}.

The origin of these corrections are particles propagating nontrivially
around a spatial dimension of the lattice. The leading contribution
obviously originates from the lightest particles, the pions. One can
more systematically treat their effect in finite volume chiral
perturbation theory
\cite{Gasser:1986vb,Gasser:1987ah,Gasser:1987zq}. The $\chi$PT treatment
has also been combined with the approach of \cite{Luscher:1985dn} in
\cite{Colangelo:2003hf,Colangelo:2005gd}. A similar expansion for
baryons has also been pioneered \cite{Colangelo:2005cg}. As one can see
from fig.~\ref{fig:l2}, current lattices are typically large enough to
have percent level or smaller finite volume corrections on the pion
mass. Note however, that corrections to baryon masses can be
substantially larger \cite{Colangelo:2005cg}.

A different sort of finite volume effect can be observed for resonant
states. In the continuum, resonances are embedded in a continuous
spectrum of scattering states. In a finite volume, the spectrum
becomes discrete and the energy of the possible scattering states
increases with shrinking volume due to the increasing magnitude of the
discrete lattice momenta. Ultimately, for small enough volumes, the
resonances end up dominating the ground state. A systematic treatment
of these finite volume energy shifts was developed in
\cite{Luscher:1986pf,Luscher:1990ux,Luscher:1991cf}. Recently it has
been suggested to use this volume, mass and momentum dependence of the
energy levels on a statistical basis and identify resonances via
eigenstate densities
\cite{Bernard:2008ax,Giudice:2010ch,Bernard:2010fp}.  An alternative
method based on finite time correlators has also recently been
suggested in \cite{Meissner:2010rq}.

Although conceptually clear, the treatment of resonant states in a
region where they are not the ground state faces the huge challenge of
reliably extracting the ground state as well as a number of excited
states. One therefore often uses the assumption that an operator which
does not mirror the valence quark structure of a scattering state will
almost exclusively couple to the resonance for extracting directly the
desired resonance level. The validity of this assumption is confirmed
by recent studies \cite{Lin:2008pr,Engel:2010my}.

Fixing the global topological charge in QCD is a restriction that
becomes irrelevant in the infinite volume limit. For this reason fixing
the topological charge in lattice QCD calculations may be viewed as
introducing an additional third type of finite volume corrections
\cite{Brower:2003yx,Aoki:2007ka}. 

\section{Ensemble overview}
\label{ens}

In order to assess currently available lattice ensembles with respect to
the three main sources of systematic error discussed in the previous
section, it is instructive to look at their position in a landscape with
respect to the four quantities: light and strange quark masses (physical
point), lattice spacing (continuum) and volume. Because light and
strange quark masses are scheme and scale dependent quantities, it is
easier to use the quantities $M_\pi$ and $\sqrt{2M_K^2-M_\pi^2}$ instead
that are proportional to the square root of the sum of light
quark masses resp. the strange quark mass to leading order.

\begin{figure}
  \includegraphics[width=\textwidth]{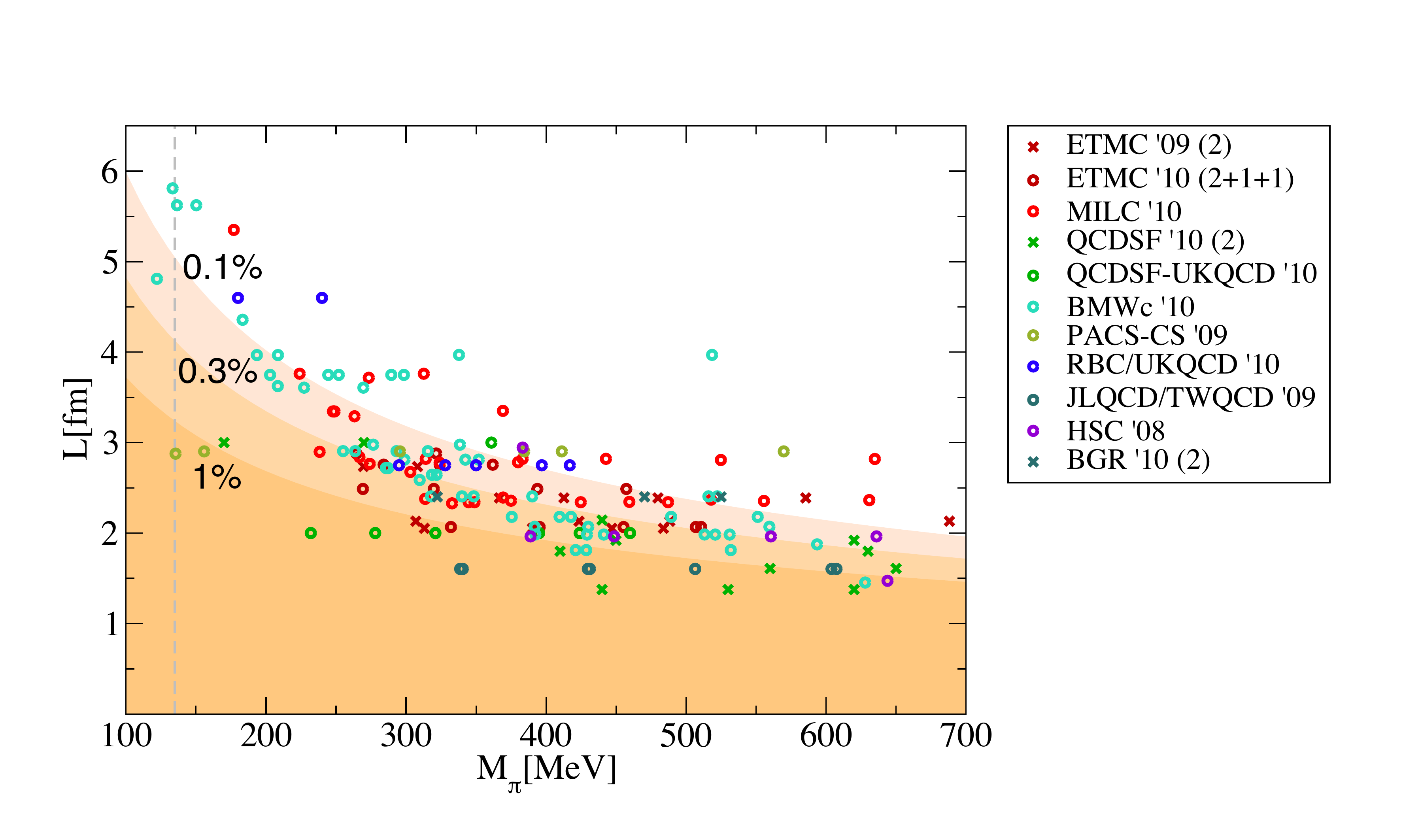} \caption{\label{fig:l2}
    The landscape of recent dynamical fermion simulations projected to
    the $L$ vs. $M_\pi$ plane.  The borders of the shaded regions are
    placed where the expected relative error of the pion mass is
    $1\%$, $0.3\%$ resp. $0.1\%$ according to
    \cite{Colangelo:2005gd}. Data points are taken from the following
    references: ETMC'09(2) \cite{Blossier:2009bx}, ETMC'10(2+1+1)
    \cite{Baron:2011sf}, MILC'10 \cite{Bazavov:2009bb}, QCDSF'10(2)
    \cite{Schierholz:2010xx}, QCDSF-UKQCD'10 \cite{Bietenholz:2010si},
    BMWc'10 \cite{Durr:2010aw,Durr:2008zz}, PACS-CS'09
    \cite{Aoki:2009ix,Aoki:2008sm}, RBC-UKQCD'10
    \cite{Aoki:2010dy,Mawhinney:2010xx}, JLQCD/TWQCD'09
    \cite{Noaki:2009sk}, HSC'10 \cite{Lin:2008pr} and BGR'10(2)
    \cite{Engel:2010my}.  All ensembles are from $N_f=2+1$ simulations
    except explicitly noted otherwise. For staggered resp. twisted
    mass ensembles, the Goldstone resp. charged pion masses are
    plotted.}
\end{figure}

\begin{figure}
\includegraphics[width=\textwidth]{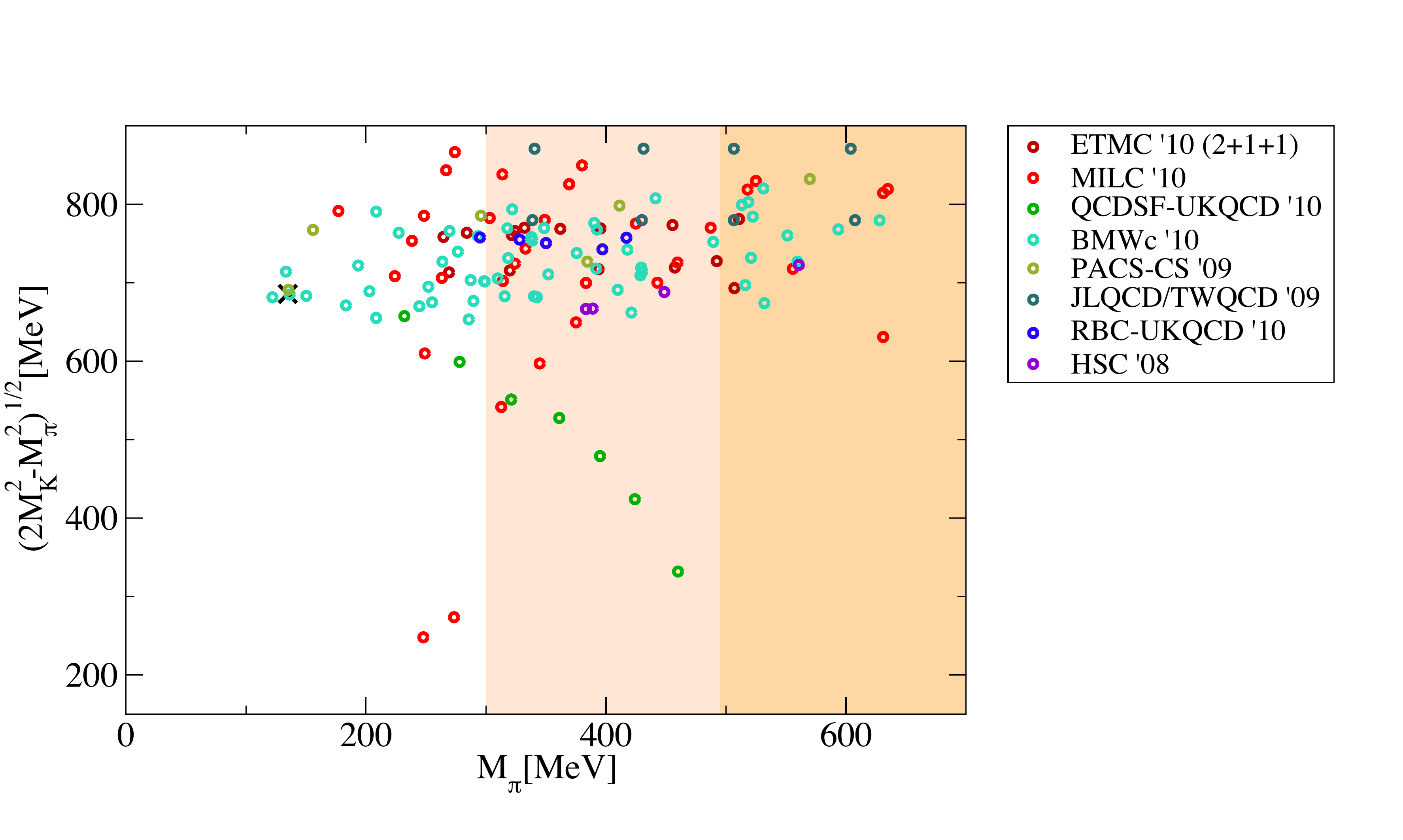}
\caption{\label{fig:l1}
The landscape of recent dynamical fermion simulations projected to the
$\sqrt{2M_K^2-M_\pi^2}$ vs. $M_\pi$ plane. The cross marks the physical
point while shaded areas with increasingly light shade indicate
physically more desirable regions of parameter space.  Data points are
taken from the following references:
ETMC'10(2+1+1) \cite{Baron:2011sf}, 
MILC'10 \cite{Bazavov:2009bb},
QCDSF-UKQCD'10 \cite{Bietenholz:2010si},
BMWc'10 \cite{Durr:2010aw,Durr:2008zz},
PACS-CS'09 \cite{Aoki:2009ix,Aoki:2008sm},
RBC-UKQCD'10 \cite{Aoki:2010dy,Mawhinney:2010xx},
JLQCD/TWQCD'09 \cite{Noaki:2009sk},
HSC'10 \cite{Lin:2008pr} and
all ensembles are from $N_f=2+1$ simulations except explicitly noted
otherwise. For staggered resp. twisted mass ensembles, the Goldstone
resp. charged pion masses are plotted.
}
\end{figure}

\begin{figure}
\includegraphics[width=\textwidth]{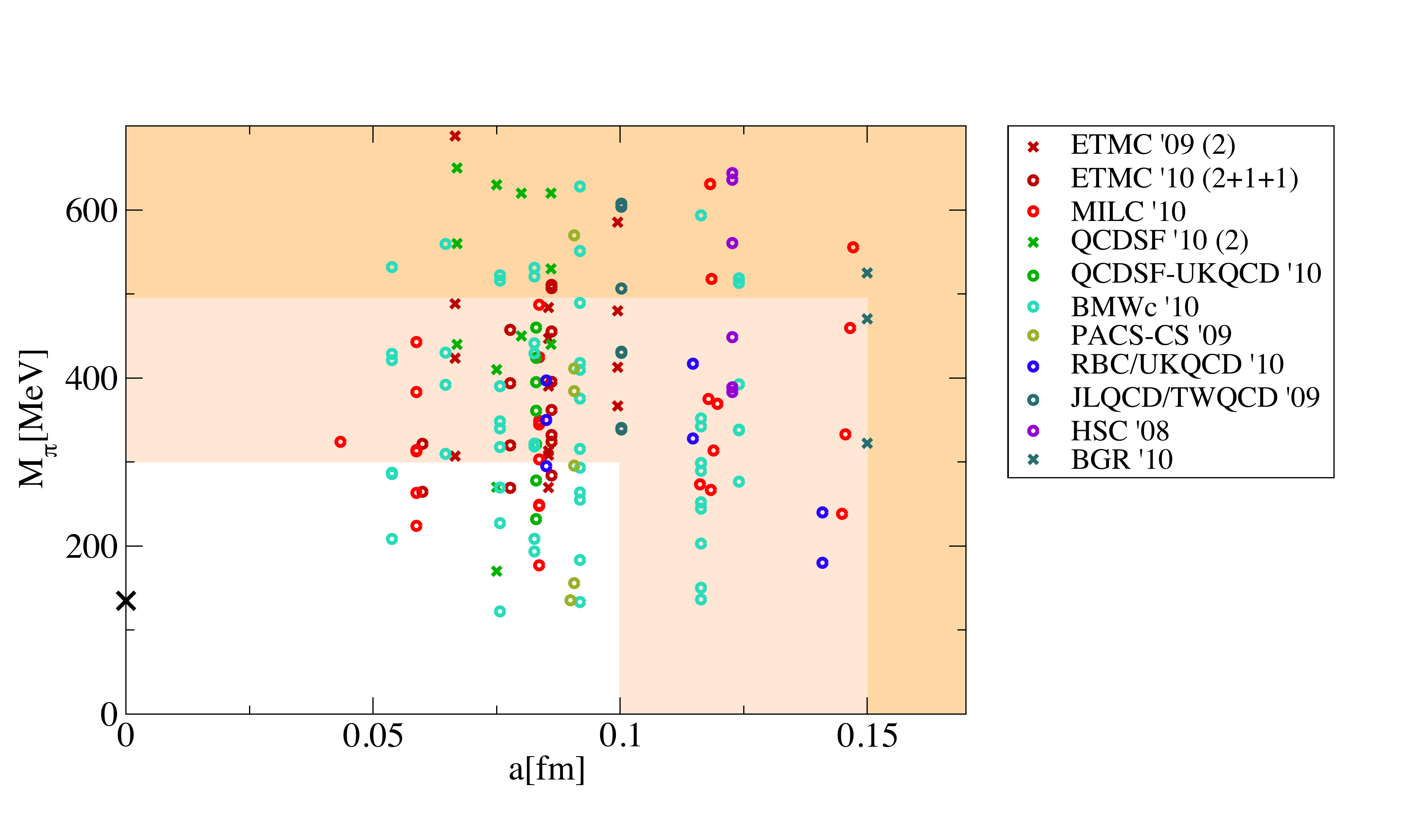}
\caption{\label{fig:l3}
The landscape of recent dynamical fermion simulations projected to the
$M_\pi$ vs. $a$ plane. The cross marks the physical point while shaded
areas with increasingly light shade indicate physically more desirable
regions of parameter space.  Data points are taken from the following
references:
ETMC'09(2) \cite{Blossier:2009bx},
ETMC'10(2+1+1) \cite{Baron:2011sf}, 
MILC'10 \cite{Bazavov:2009bb},
QCDSF'10(2) \cite{Schierholz:2010xx},
QCDSF-UKQCD'10 \cite{Bietenholz:2010si},
BMWc'10 \cite{Durr:2010aw,Durr:2008zz},
PACS-CS'09 \cite{Aoki:2009ix,Aoki:2008sm},
RBC-UKQCD'10 \cite{Aoki:2010dy,Mawhinney:2010xx},
JLQCD/TWQCD'09 \cite{Noaki:2009sk},
HSC'10 \cite{Lin:2008pr} and
BGR'10(2) \cite{Engel:2010my}.
All ensembles are from $N_f=2+1$ simulations except explicitly noted
otherwise. For staggered resp. twisted mass ensembles, the Goldstone
resp. charged pion masses are plotted.
}
\end{figure}

In figs.~\ref{fig:l1}-\ref{fig:l3} three projections of this landscape
are plotted. The first one, fig.~\ref{fig:l1}, displays the position of
current ensembles in the $\sqrt{2M_K^2-M_\pi^2}$ vs. $M_\pi$ plane. As one
can see, the physical point has already been reached. In
fig.~\ref{fig:l2} the landscape is projected to the $L$ vs. $M_\pi$
plane. One observes that the bulk of current day lattice ensembles lies
in a region where the pion mass is expected to be affected by finite
volume corrections by less than one per cent. Finally, fig.~\ref{fig:l3}
displays a projection to the $M_\pi$ vs. $a$ plane. Also here one can
see that present day simulations start populating the interesting
regime of physical or near physical pion masses at a range of relatively
small lattice spacings.

The important point is of course to have ensembles that simultaneously
lie in the desirable regions with respect to all coordinates of the
landscape, i.e. that are at or close to the physical point at large
volumes and a range of relatively small lattice spacings.

\section{Ground states}
\label{gs}

\begin{figure}
\centerline{\includegraphics[width=0.7\textwidth]{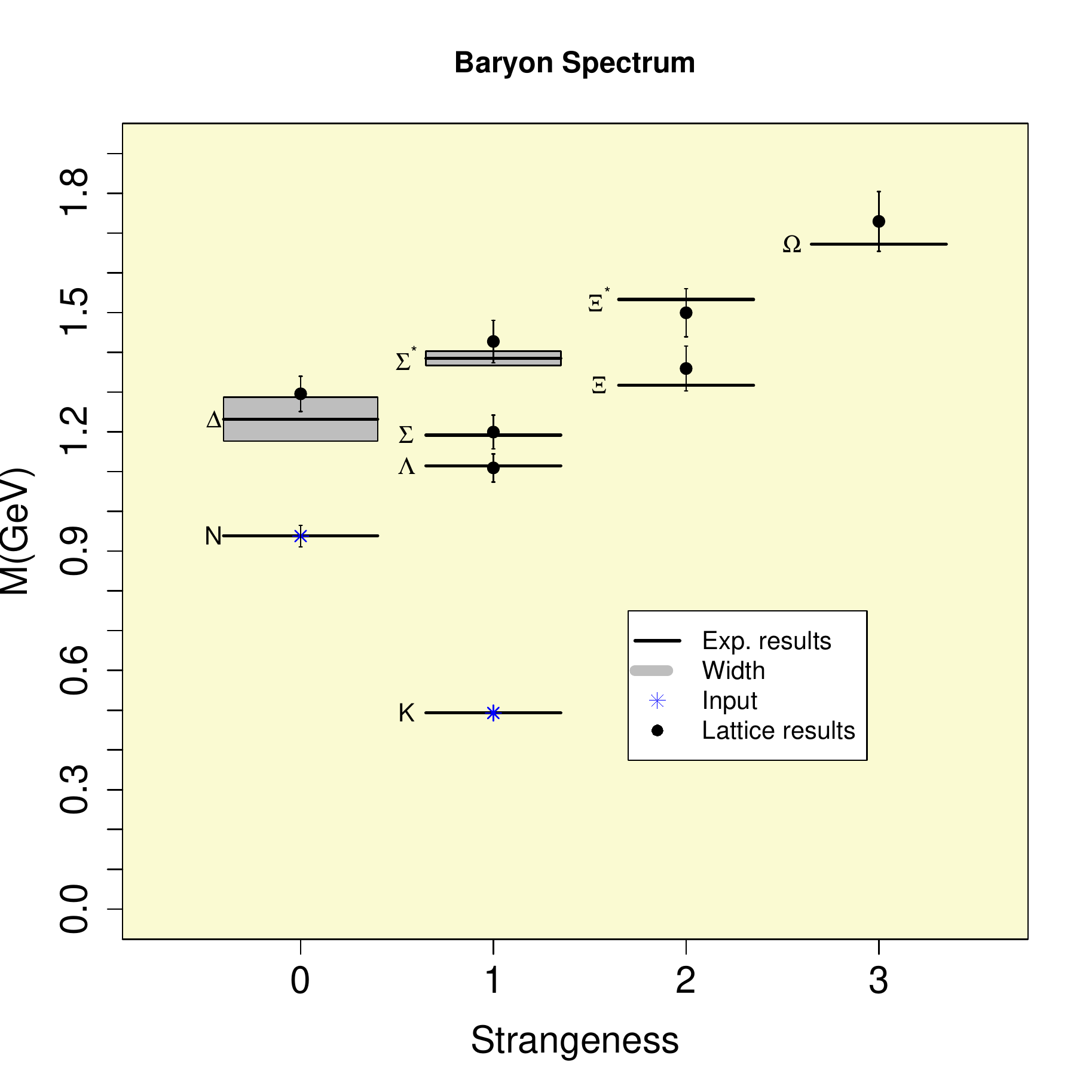}}
\caption{\label{fig:etm2}
Baryon spectrum obtained by the ETM
collaboration with $N_f=2$ twisted mass fermions. The plot is reproduced
from \cite{Alexandrou:2009qu} with friendly permission of the ETM
collaboration.
}
\end{figure}

We begin our overview of calculations of the light hadron spectrum by
results of the ETM collaboration for $N_f=2$ twisted mass fermions
\cite{Alexandrou:2009qu}. The ETM collaboration used $N_f=2$ twisted
mass fermions at maximal renormalized twist on a tree level Symanzik
improved gauge action. Two lattice spacings ($a\sim 0.07\text{~fm}$
and $a\sim 0.09\text{~fm}$) were used with charged pion masses in the
range $270$ to $500\text{~MeV}$.\footnote{The isospin splitting of the pions is
${M^\pm_\pi}^2-{M^0_\pi}^2\sim (150-220\text{~MeV})^2$
\cite{Baron:2009wt}} The lattice spacing was set via the nucleon mass
and chiral extrapolations were performed with a variety of different
ans\"atze. The valence strange quark mass is set by tuning the kaon
mass to its physical value. The final result, displayed in
fig.~\ref{fig:etm2}, employs $O(p^3)$ resp. NLO SU(2) heavy barion
$\chi$PT chiral extrapolation for baryons without resp. with valence
strange quarks. The continuum extrapolation was performed using a
constant which was demonstrated to be sufficient at the given level of
accuracy.  Exponential finite volume corrections were taken into
account in the final fit form. Resonant state finite volume
corrections were not performed but are believed to be irrelevant in
the region of parameter space covered by the simulations. Effects of
the twisted mass isospin breaking were observed to be negligible
except in the case of the $\Xi$ where they amounted to a $6\%$
correction.

The current results on light hadron masses from the MILC collaboration
are summarized in \cite{Bazavov:2009bb}. For the baryon mass analysis
ensembles at 3 of the currently available 6 lattice spacings $a\sim
0.06\text{~fm}$, $a\sim 0.09\text{~fm}$ and $a\sim 0.12\text{~fm}$ are
used. The smallest Goldstone pion mass is $\sim 180\text{~MeV}$
corresponding to an RMS pion mass of $\sim 250\text{~MeV}$. Both gauge
and fermion (asqtad) actions have $O(\alpha_s a^2)$ scaling behavior and
the continuum extrapolation is done linearly in this quantity. Depending
on the specific observable, chiral or polynomial fit forms are found to
best describe the data and the scale is set via $r_1$. The final
results, presented in \cite{Bazavov:2009bb} are in good agreement with
the observed hadron spectrum.

A subset of the MILC ensembles with $a\sim 0.12\text{~fm}$ and a smallest
pion mass of $\sim 290 \text{~MeV}$ has been studied in
\cite{WalkerLoud:2008bp} in a mixed action setup with domain wall
valence quarks. Comparing different chiral fit forms for the nucleon
mass it was demonstrated that a simple linear fit in $M_\pi$ gives the
best description of the data and extrapolates to the correct value at
the physical point. In the same paper, this feature has also been found
in other collaborations data. 

The QCDSF-UKQCD collaboration has recently proposed a different approach to
the physical point starting from an SU(3) symmetric theory and
systematically expanding in the SU(3) breaking parameter while keeping
$2M_K^2+M_\pi^2$ constant \cite{Bietenholz:2010jr,Bietenholz:2010si}.
Preliminary results at a single lattice spacing show a linear dependence
of the octet and decuplet masses considered and a good agreement with
the experimentally observed hadron spectrum. An $N_f=2+1$ nonperturbatively
improved single step stout smeared clover action on a tree level
Symanzik improved gauge action was used for this study. Finite size
corrections are not yet applied.

\begin{figure}
\centerline{\includegraphics[width=0.87\textwidth]{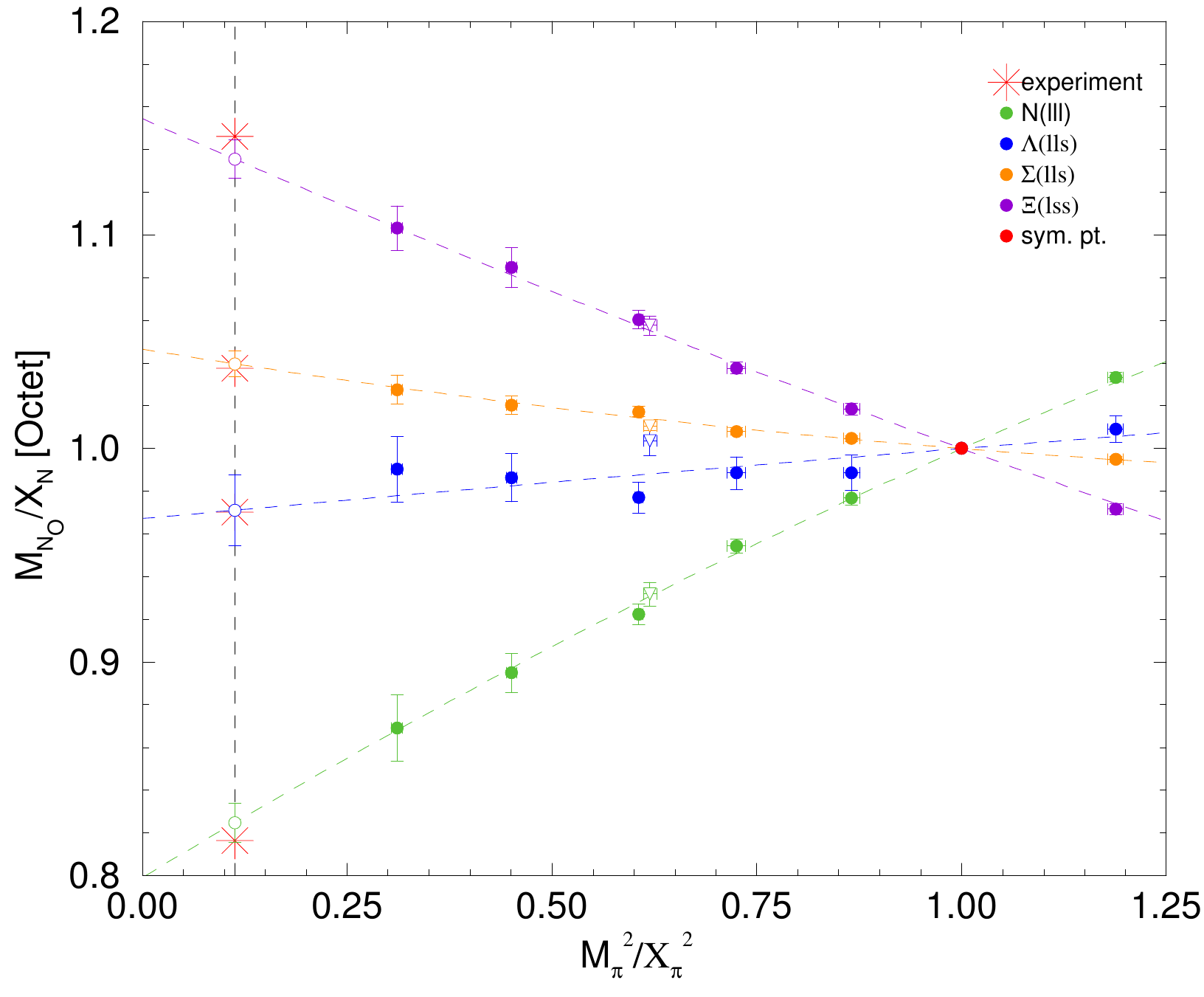}}
\caption{\label{fig:qcdsf}
Chiral behavior of the ratio of individual octet masses over the average
octet mass $X_N=\frac{1}{3}\left(M_N+M_\Sigma+M_\Xi\right)$ vs. the
ratio of the square of the pion mass over the square average of
pseudoscalar meson masses
 $X_\pi^2=\frac{1}{3}\left(2M^2_K+M^2_\pi\right)$ as obtained by the
 QCDSF-UKQCD collaboration. The plot is reproduced from
 \cite{Bietenholz:2010si} with friendly permission of the QCDSF-UKQCD collaboration.}
\end{figure}

The light hadron spectrum calculation of the
Budapest-Marseille-Wuppertal collaboration (BMWc)
\cite{Durr:2008zz} was performed with tree level improved 6-step stout
smeared $N_f=2+1$ clover fermions on a tree level Symanzik improved
gauge action. Pion masses down to $190\text{~MeV}$ and three lattice
spacings $a\sim0.06\text{~fm}$, $a\sim0.8\text{~fm}$ and
$a\sim0.12\text{~fm}$ were used. Finite volume corrections were applied
including energy shifts for resonant states. The continuum extrapolation
was performed with a term linear in $a$ or $a^2$ and chiral fits were
done with both Taylor and NLO heavy baryon $\chi$PT with a free coefficient.
The systematic error was estimated by the spread of the results of all
analyses weighted by the fit quality. Good agreement with the
experimentally observed light hadron spectrum was found.

\begin{figure}
\centerline{\includegraphics[width=0.87\textwidth]{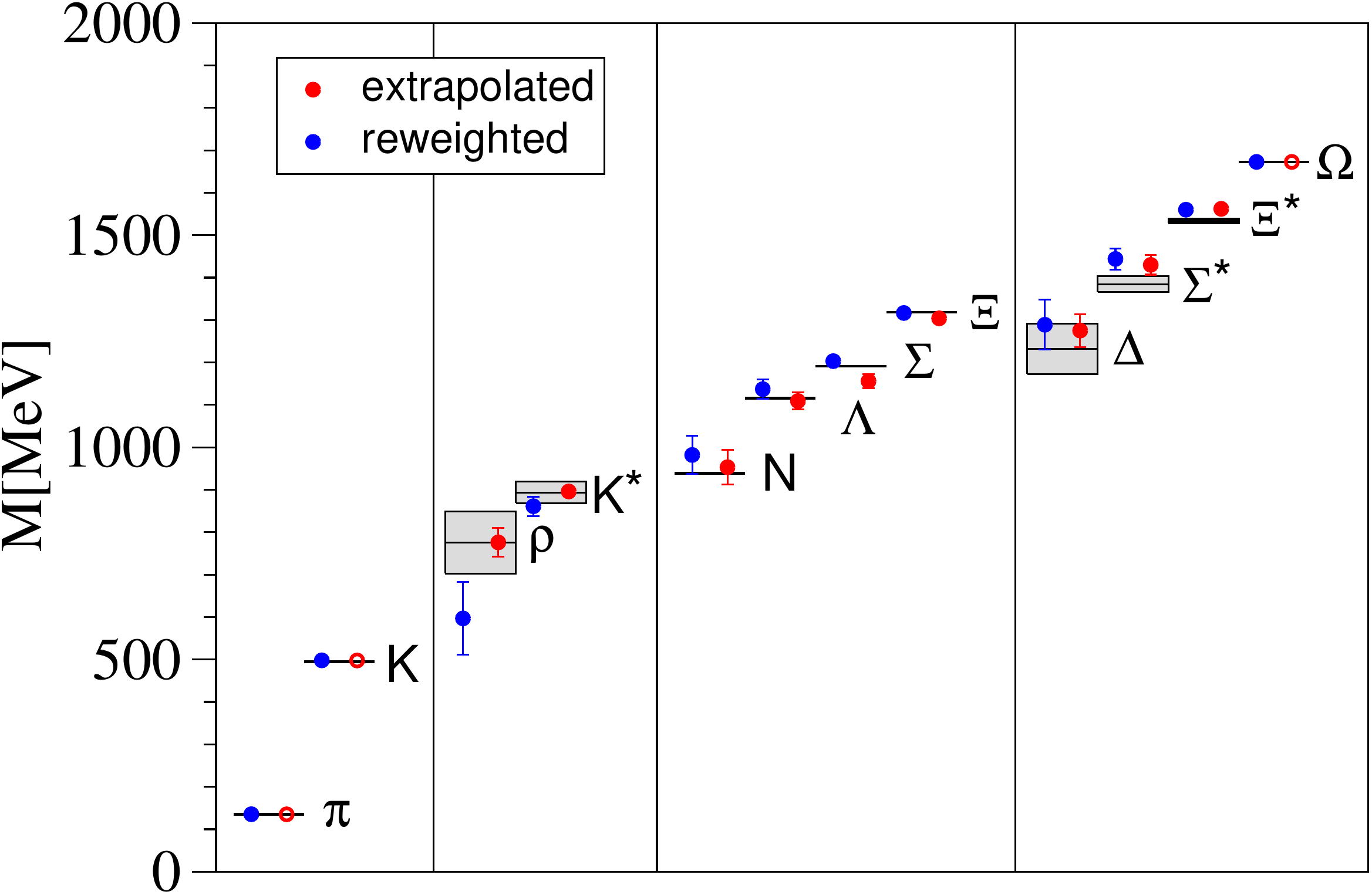}}
\caption{\label{fig:pacscs}
Comparison of the light hadron spectrum results from PACS-CS extrapolated
\cite{Aoki:2008sm} and reweighted \cite{Aoki:2009ix} ensembles with
 experiment.
}
\end{figure}

The PACS-CS collaboration has published results for the light hadron
spectrum using both a chiral extrapolation \cite{Aoki:2008sm} and a
direct reweighting to the physical point \cite{Aoki:2009ix}. In both
cases $N_f=2+1$ nonperturbatively $O(a)$ improved cover fermions on an
Iwasaki gauge action were used at a single lattice spacing $a\sim
0.09\text{~fm}$. Pion masses down to $\sim 150\text{~MeV}$ were directly
simulated and a reweighting to the physical point was carried out with
the lightest ensemble. In the extrapolated ensemble finite size effects
on the pseudoscalar masses were corrected using SU(2) $\chi$PT at
NLO. The tiny chiral extrapolation was performed linearly in the light
quark mass and $M_\Omega$ was used to set the scale. More involved
chiral forms were subsequently investigated in \cite{Ishikawa:2009vc}.
Similarly in the reweighted ensemble the masses of the $\pi$, $K$ and
$\Omega$ were used to tune to the physical point. The resulting vector
meson and baryon spectra are in good agreement with experiment (see
fig.~\ref{fig:pacscs}).

The RBC-UKQCD collaboration has recently performed a pioneering
calculation of the $\eta$ and $\eta^\prime$ masses using $N_f=2+1$
flavor domain wall ensembles on an Iwasaki gauge action
\cite{Christ:2010dd}. Three pion masses in the range $400-700\text{~MeV}$
on a single lattice spacing $a\sim 0.11\text{~fm}$ were used. A two
operator basis with gauge fixed wall sources was used to extract the
correlation functions. A mixing angle of $\Theta=-9.2(4.7)^\circ$ and
masses $M_\eta=583(15)\text{~MeV}$ and
$M_{\eta^\prime}=853(123)\text{~MeV}$ were found.

\section{Electromagnetic effects}
\label{em}

As the precision in lattice hadron spectroscopy is increasing,
subleading effects such as isospin splitting and QED corrections start
to become relevant for certain observables.

Electromagnetic splitting in the hadron spectrum was recently
investigated in quenched, noncompact QED \cite{Blum:2010ym} on RBC/UKQCD
$N_f=2+1$ domain wall ensembles with Iwasaki gauge action at
$a\sim0.11\text{~fm}$. In a partially quenched setup valence pions down
to a mass of $\sim 250\text{~MeV}$ were used. Chiral fits were performed
using SU(2) and SU(3) $\chi$PT at NLO. Systematic error estimates due to
the chiral extrapolation, finite lattice spacing, finite volume and QED
quenching were presented. The electromagnetic hadron mass splittings
found are summarized in table~\ref{tab:em}.

\begin{table}
\begin{tabular}{l||c|c|c}
& $\left(M_{\pi^+}-M_{\pi^0}\right)_{\text{QED}}$  &
 $\left(M_{K^+}-M_{K^0}\right)_{\text{QED}}$ &
 $\left(M_{n}-M_{p}\right)_{\text{QED}}$ \\
\hline
\cite{Blum:2010ym} & 4.50(23) MeV& 1.33(4) MeV& -0.38(7) MeV\\
\cite{Portelli:2010yn} & 5.1(1.1) MeV& 2.2(0.2) MeV& -
\end{tabular}
\caption{
\label{tab:em} Comparison of quenched QED results for the electromagnetic
 splitting of hadron masses from \cite{Blum:2010ym} and
 \cite{Portelli:2010yn}. All errors are statistical only.
}
\end{table}

At this conference, the Budapest-Marseille-Wuppertal collaboration has
also presented preliminary results of electromagnetic meson mass
splitting \cite{Portelli:2010yn} using quenched, noncompact QED on a
subset of their $N_f=2+1$, 2 step HEX smeared clover ensembles on a
tree level Symanzik improved gauge action. Four ensembles with pion
masses in the range $200-450\text{~MeV}$ were analyzed and an extra
additive mass renormalization was applied to light valence quarks in
order to retune the theory to an almost unitary point with degenerate
$u$ and $d$ masses. In addition a constant electromagnetic background
field was introduced to cancel leading finite volume effects. First
preliminary results are given in table~\ref{tab:em}.

Progress has also been reported at this conference regarding the
inclusion of quenched QED effects into rooted staggered $\chi$PT
\cite{Bernard:2010qd}. This calculation will complement forthcoming
MILC simulations including quenched QED that are currently in progress.

\section{Excited states}
\label{es}

Hadron spectroscopy of excited states is a substantially more difficult
task than it is for ground states. While the extraction of ground
state masses (at least in the case of non-resonant states with no
disconnected contributions) is relatively straightforward, the
extraction of higher excited states is substantially more
challenging. Excited state contributions to the propagator are
exponentially suppressed at large Euclidean times and the relevant
energy levels may be hidden between a number of scattering states.
Accordingly, finite volume corrections or a continuum extrapolation that
are common in ground state spectroscopy are currently beyond the level
of precision obtainable in excited state spectroscopy. In some cases
even a controlled chiral extrapolation or $N_f=2+1$ dynamical
quarks are currently beyond reach.

\begin{figure}
\includegraphics[width=0.33\textwidth]{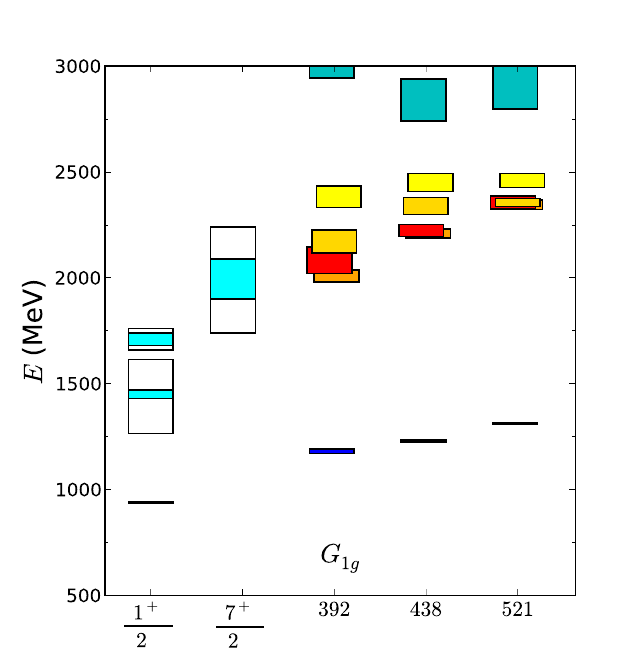}
\includegraphics[width=0.33\textwidth]{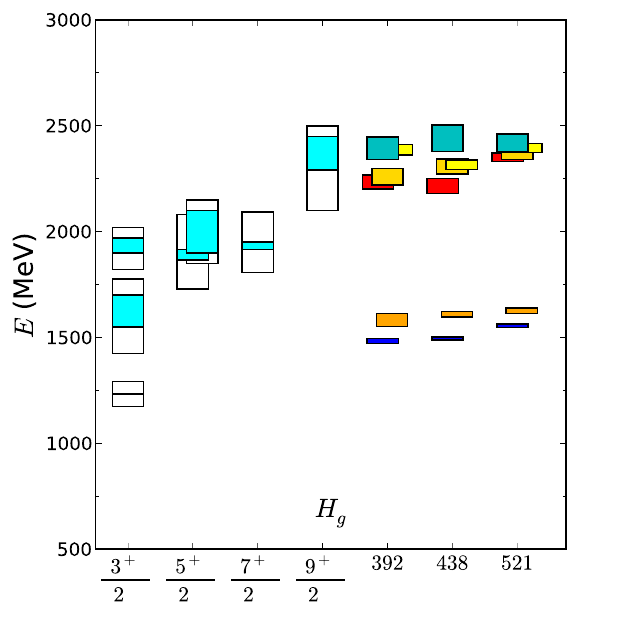}
\includegraphics[width=0.33\textwidth]{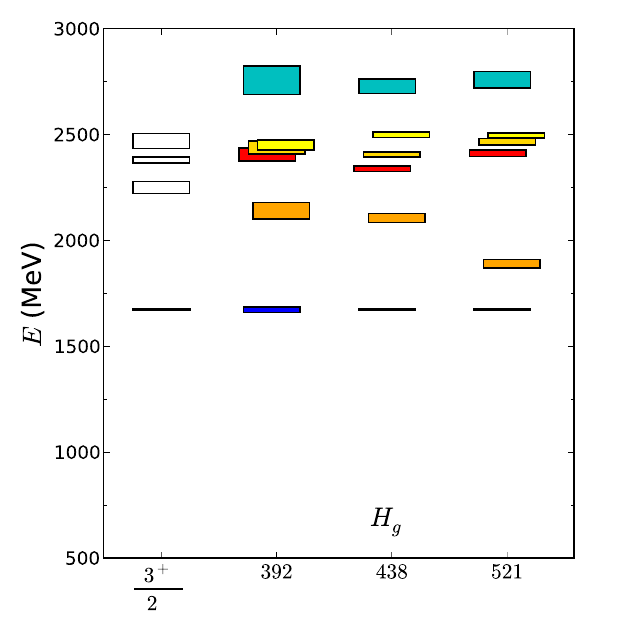}
\caption{\label{fig:hsc}
Comparison of part of the excited state spectrum of nucleon (left)
$\Delta$ (middle) and $\Omega$ (right) type baryons as computed by the hadron
spectrum collaboration at three different pion masses with
experiment. More details can be found in the original paper.  The plot
is reproduced from \cite{Bulava:2010yg} with friendly permission of the
hadron spectrum collaboration.
}
\end{figure}

The hadron spectrum collaboration is using anisotropic lattices in order
to obtain a fine time resolution of the propagators. The lattice spacing
in time direction is tuned to be smaller by a factor of $\xi\sim 3.5$
than the lattice spacing in the spatial directions
\cite{Edwards:2008ja}. In their excited state spectroscopy studies
\cite{Lin:2008pr,Bulava:2010yg,Bulava:2010vk} they employ $N_f=2+1$
anisotropic clover fermions on a tree level tadpole improved Symanzik
gauge action. A single spatial lattice spacing $a_s\sim 0.12\text{~fm}$
and three pion masses in the range $390-530\text{~MeV}$ are used. The
scale is set with $M_\Omega$. A variational method based on a large
number (6-10) of specifically tailored interpolating operators are used
to extract the tower of excited states in the different
channels. Results are reported at three different pion masses and show a
nice overall qualitative agreement with the experimentally observed
excited hadron spectrum (see fig.~\ref{fig:hsc}). The authors emphasize
the need for multi hadron interpolating operators in order to reliably
identify scattering states.

On the more conceptual side, progress in constructing optimized single
hadron operators via a combination of Laplacian-Heavyside (LapH) smeared
sources with stochastic estimator techniques was presented in
\cite{Foley:2010vv} and first results of this technique on anisotropic
HSC lattices have been reported in \cite{Morningstar:2010kb}.

The BGR collaboration has computed ground and excited state hadron
spectra using $N_f=2$ single step stout smeared chirally improved
fermions on a tadpole improved L\"uscher-Weisz gauge action at a singe
lattice spacing $a\sim 0.15\text{~fm}$
\cite{Engel:2010my,Engel:2009cq,Engel:2010dy}. Three pion masses in the
range $320-530\text{~MeV}$ were used and the scale was set with $r_0$
. Gaussian smeared quark sources were used in combination with a
variational method based on three interpolating operators to extract the
energy levels. A chiral extrapolation linear in $M_\pi$ was performed
and the strange quark was introduced in a partially quenched setup. A
good signal for the ground state was found but excited and scattering
state signals were generally weak.

A long standing puzzle in excited state baryon spectroscopy is the
correct ordering of the negative parity ground state $N^{\frac{1}{2}-}$
and the first positive parity excitation of the nucleon (the Roper
resonance). Experimentally, the Roper state has a mass of
$1440\text{~MeV}$ while the mass of the $N^{\frac{1}{2}-}$ is
$1553\text{~MeV}$. On the lattice however, the Roper is usually found to be
heavier than the $N^{\frac{1}{2}-}$.

The CSSM collaboration has recently proposed a method of analyzing
eigenstate projected correlation functions of a full multi-source
propagator matrix \cite{Mahbub:2009nr}. The method was applied to
investigate the level ordering in the excited nucleon system both on
quenched \cite{Mahbub:2009aa,Mahbub:2010jz,Mahbub:2010vu} and $N_f=2+1$
dynamical \cite{Mahbub:2010me,Mahbub:2010rm} configurations. Fat link
irrelevant clover (FLIC) valence fermions were used on either a quenched
DBW2 gauge action at a single lattice spacing $a\sim 0.13\text{~fm}$ that
was determined by $r_0$ or the PACS-CS dynamical ensembles discussed in
sect.~\ref{gs}. Large operator bases of up to 8 were used and signals
for up to 3 excited states were identified. The chiral behavior of both
positive and negative nucleon excitations was studied and some evidence
was found for the correct ordering of the negative parity ground state
and the Roper resonance as one approaches physical pion masses.

\section{Decay constants $f_\pi$ and $f_K$}
\label{fp}

We now turn our attention to recent results for the pseudoscalar decay
constants $f_\pi$ and $f_K$.

The ETM collaboration has performed an analysis of decay constants in
$N_f=2$ twisted mass QCD \cite{Blossier:2009bx,Baron:2009wt} where
$f_\pi$ was used to set the scale. The analysis in
\cite{Blossier:2009bx} is based on an extension of the ensembles used in
the spectrum calculation \cite{Alexandrou:2009qu} with a third, coarser
lattice spacing $a\sim 0.10\text{~fm}$. The strange quark is introduced
in a partially quenched setup. A combined chiral and continuum
extrapolation is performed and finite size effects have been corrected
for. The continuum extrapolation has been performed with $O(a^2)$ and
$O(a^2\mu)$ terms (where $\mu$ is the twisted light mass) while the
chiral fit was done using SU(2) $\chi$PT at NLO with an SU(3) fit
providing a crosscheck. The final result is given as
$f_K/f_\pi=1.210(18)$.

At this conference, the ETM collaboration has also presented first
preliminary results for the decay constants in $N_f=2+1+1$ twisted mass
QCD \cite{Baron:2011sf,Farchioni:2010tb}. They used Iwasaki gauge and a
mixed action setup with Osterwalder-Seiler valence quarks. A crosscheck
was performed in a unitary setup and unitarity violating effects were
found to be within the statistical error. The two finest of the
available three lattice spacings $a\sim 0.08\text{~fm}$ and $a\sim
0.06\text{~fm}$ were used and a constant continuum extrapolation was
performed. Finite size effects were corrected using $\chi$PT and
$f_\pi$ was again used to set the scale. The preliminary result with
statistical errors only is $f_K/f_\pi=1.224(13)$. Results for individual
decay constants have not been given but preliminary results on the axial
renormalization constants are available \cite{Dimopoulos:2011wz}.

\begin{figure}
\centerline{\includegraphics[width=0.87\textwidth]{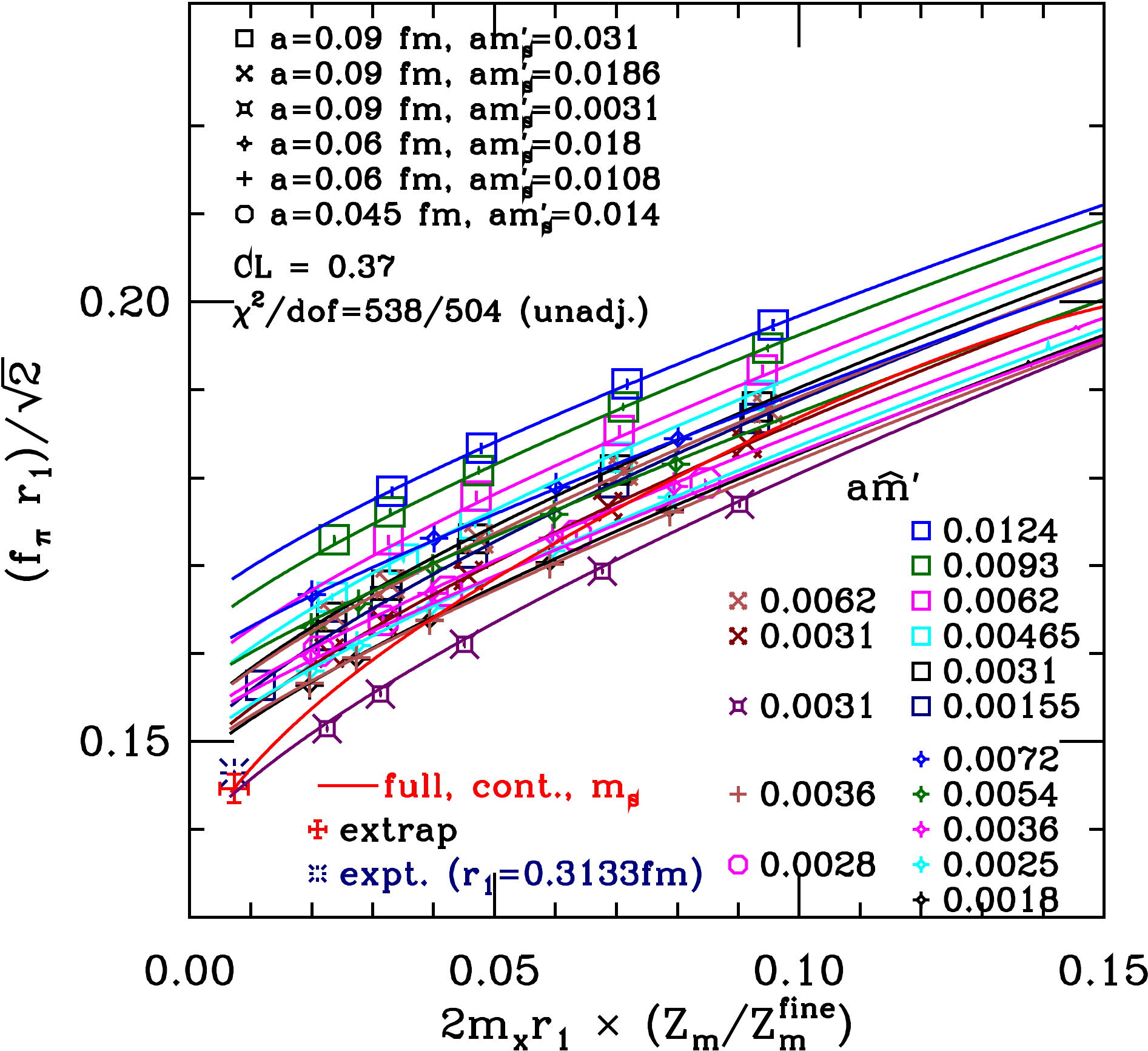}}
\caption{\label{fig:MILC}
The MILC collaborations SU(3) chiral fits to partially quenched lattice
data of the pseudoscalar decay constant in terms of the scale setting
parameter $r_1$. The figure displays part of a combined fit to decay
constants and quark masses. Reproduced from \cite{Bazavov:2010hj} with
friendly permission of the MILC collaboration.}
\end{figure}

At this conference, the MILC collaboration has presented new preliminary
results on pseudoscalar decay constants using their 3 finest sets of
partially quenched $N_f=2+1$ asqtad rooted staggered fermions
\cite{Bazavov:2010hj} updating a previous analysis
\cite{Bazavov:2009tw,Bazavov:2009fk}. They use an elaborate combined fit
function for both pseudoscalar decay constants and pseudoscalar masses
incorporating full NLO staggered $\chi$PT and the full NNLO continuum
expression. The LECs of this SU(3) chiral form are first fit in a region
of artificially small strange quark mass to avoid convergence issues
with the SU(3) chiral expansion. Finally, N$^3$LO and N$^4$LO analytic
terms are added and a fit to the large ensemble with approximately
physical strange quark mass is performed. This final fit has 504 degrees
of freedom and its decay constant part is displayed in
fig.~\ref{fig:MILC}. Using the HPQCD value of $r_1$
\cite{Davies:2009tsa} one obtains $f_\pi=129.2(0.4)(1.4)\text{~MeV}$, which
is in nice agreement with the value obtained by an SU(2) chiral fit
$f_\pi=128.3(9)\left({+2.0\atop -8}\right)\text{~MeV}$ \cite{Bazavov:2009ir}. With
physical $f_\pi$ input they further find
$f_K/f_\pi=1.197(2)\left({+3\atop -7}\right)$.

Due to an updated scale determination, the HPQCD collaboration has given
an update \cite{Davies:2010ip} on their older values of $f_K$ and
$f_\pi$ from \cite{Follana:2007uv}. The new results are
$f_\pi=132(2)\text{~MeV}$ $f_K=159(2)\text{~MeV}$. The decay constant
ratio has not been updated and remains at
$f_K/f_\pi=1.189(2)(7)$.

The JLQCD/TWQCD collaboration has published a first preliminary value of
$f_K/f_\pi$ with $N_f=2+1$ dynamical overlap quarks on Iwasaki gauge
action in a fixed topological sector \cite{Noaki:2009sk}. Ensembles with
a combination of 5 different pion masses in the range
$340-870\text{~MeV}$ and two different strange quark masse were produced
at the single lattice spacing $a\sim 0.1\text{~fm}$. Chiral fits were
performed using NNLO SU(3) $\chi$PT and the scale was set with the
physical value of $f_\pi$. The results were corrected for conventional
finite size effects as well as those associated with fixing the
topology. The preliminary result obtained is $f_K/f_\pi=1.210(12)$ where
the errors are statistical only.

The QCDSF-UKQCD collaboration has presented at this conference a first
determination of the pseudoscalar decay constant ratio on their
$N_f=2+1$ ensembles \cite{Horsley:2010xx}. The preliminary result is
$f_K/f_\pi=1.221(15)$.

\begin{figure}
\centerline{\includegraphics[width=0.87\textwidth]{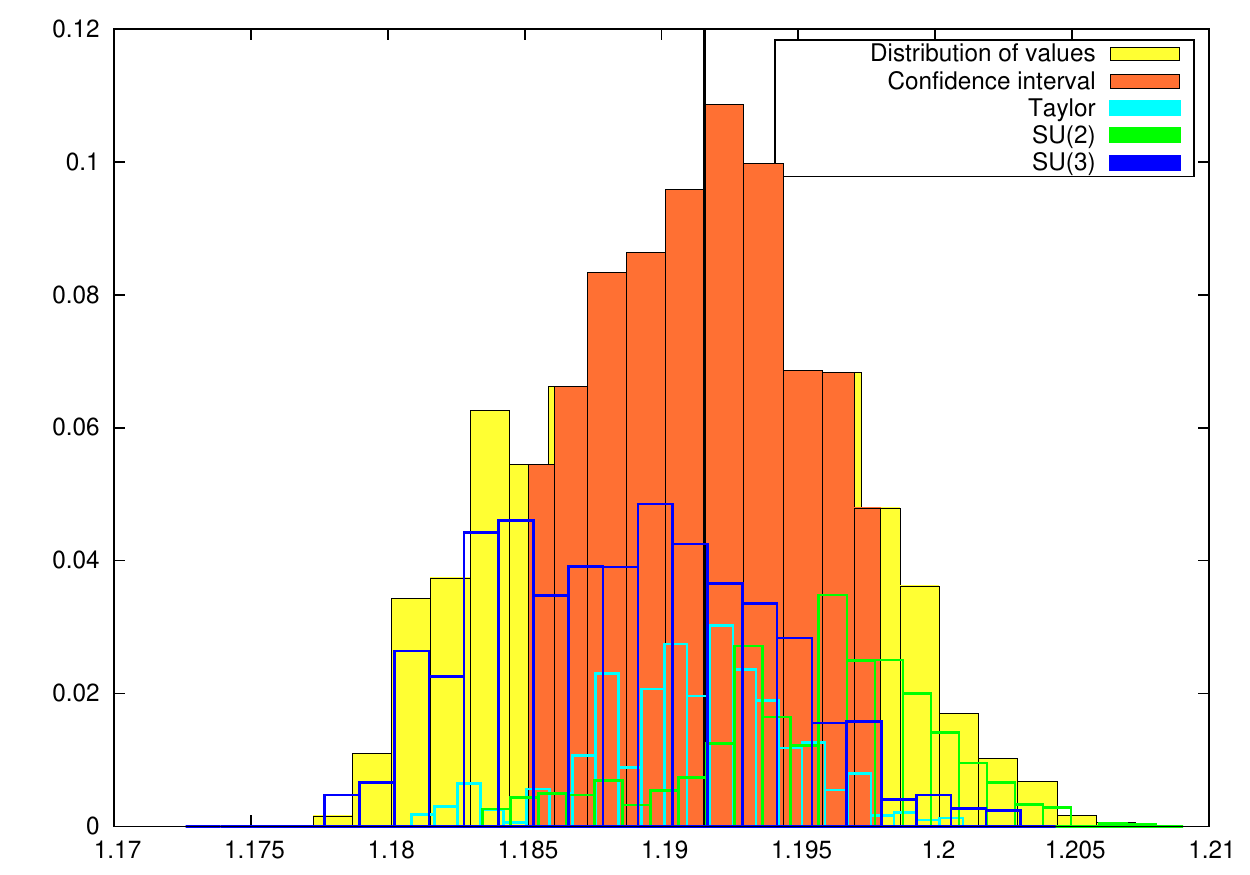}}
\caption{\label{fig:bmw} 
The histogram of results for $f_K/f_\pi$ from \cite{Durr:2010hr} over
different analysis procedures weighted by their respective fit
 quality. Subhistograms for the three different chiral fit forms are
 also plotted. One can see that their spread is smaller than their width
 and their integrated fit quality is approximately equal. 
}
\end{figure}

Using the same ensembles as for the light hadron
spectrum\cite{Durr:2008zz}, the BMWc collaboration has recently
determined the pseudoscalar decay constant ratio \cite{Durr:2010hr}.
Finite volume corrections were applied and the continuum extrapolation
was done using either $a$ or $a^2$ terms. In addition three different
chiral fit forms (SU(2) resp. SU(3) $\chi$PT at NLO and a Taylor
series expansion) were used and found to be in perfect agreement as
shown in fig.~\ref{fig:bmw}. The systematic error was estimated by the
spread of the results yielding the final result
$f_K/f_\pi=1.192(7)(6)$.

The PACS-CS collaboration has published preliminary values of the
pseudoscalar decay constants both on their $a\sim 0.09\text{~fm}$
unreweighted \cite{Aoki:2008sm} and reweighted \cite{Aoki:2009ix}
ensembles. For the extrapolated ensemble, SU(2) and SU(3) $\chi$PT
chiral forms were used and finite size effects were treated. For the
extrapolated ensemble, the renormalization constant $Z_A$ was computed
in 1-loop perturbation theory. For the reweighted ensemble $Z_A$ was
determined nonperturbatively in a Schr\"odinger functional method
\cite{Aoki:2010wm}. In both cases, the scale was set by $M_\Omega$.  The
results are $f_\pi=134.0(4.2)\text{~MeV}$, $f_K=159.4(3.1)\text{~MeV}$ and
$f_K/f_\pi=1.189(20)$ for the extrapolated and
$f_\pi=124.1(8.5)(0.8)\text{~MeV}$, $f_K=165.5(3.4)(1.0)\text{~MeV}$ and
$f_K/f_\pi=1.333(72)$ for the reweighted ensemble where the first error
is statistical and the second due to renormalization.

The RBC-UKQCD collaboration has recently published final results of
pseudoscalar decay constants using $N_f=2+1$ partially quenched domain
wall quarks on an Iwasaki gauge action \cite{Aoki:2010dy}. Ensembles
were produced at two lattice spacings $a\sim 0.08\text{~fm}$ and $a\sim
0.11\text{~fm}$ and a continuum extrapolation linear in $a^2$ was
performed. The scale was set using $M_\Omega$ and ensembles were
reweighted to the physical strange mass. Analytic and NLO SU(2) $\chi$PT
forms were used to describe the chiral behavior with finite volume
corrections were taken into account. The renormalization constant $Z_A$
was computed nonperturbatively using ratios of local and conserved
current matrix elements. The final results obtained are
$f_\pi=124(2)(5)\text{~MeV}$, $f_K=149(2)(3)\text{~MeV}$ and
$f_K/f_\pi=1.204(7)(25)$. The authors report a marked difference in the
continuum extrapolated value of $f_\pi$ depending on the fit form used
with NLO SU(2) $\chi$PT giving significantly lower results than a
leading order Taylor fit.

\begin{figure}
\centerline{\includegraphics[width=0.87\textwidth]{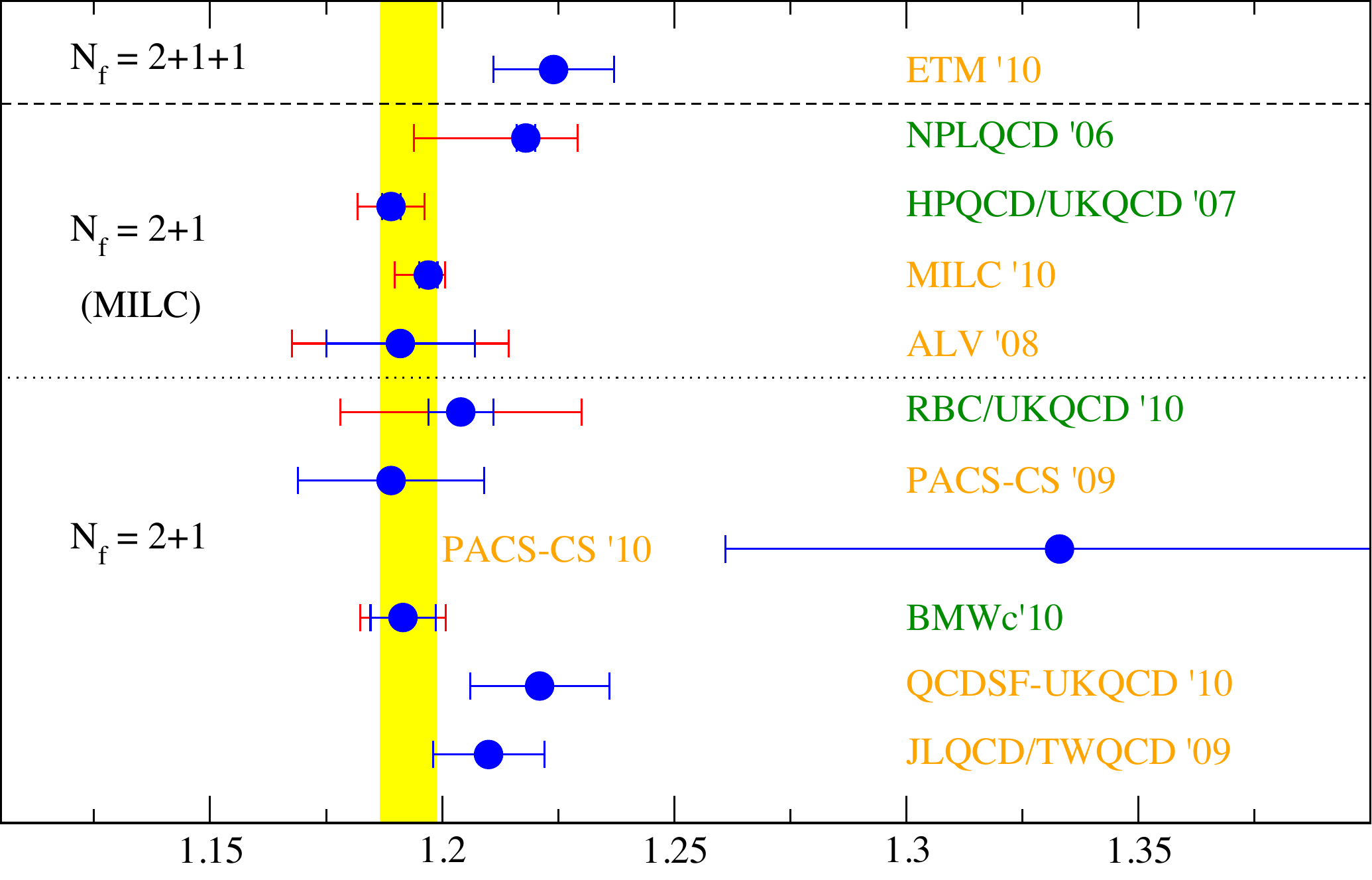}}
 \caption{
\label{fig:fkfpi}
Comparison of recent results for the ratio of decay constants
$f_K/f_\pi$ in $N_f=2+1$ and $N_f=2+1+1$ QCD.  Green entries correspond
to final results while orange ones are preliminary. The four
determinations based on the MILC ensembles are separated in the plot.
Statistical (blue) and full (red) error bars are given where
available. The yellow band corresponds to an indirect determination
assuming first row CKM unitarity and using the values
$|V_{ud}|=0.97425(22)$ from \cite{Hardy:2008gy}, $|V_{ub}|=0.00389(44)$ from
\cite{Nakamura:2010zzi} and
$\frac{|V_{us}|}{|V_{ud}|}\frac{f_K}{f_\pi}=0.27599(59)$ from
\cite{Antonelli:2008jg}. Lattice results are taken from
ETM'10 \cite{Farchioni:2010tb},
NPLQCD'06 \cite{Beane:2006kx},
HPQCD/UKQCD'07 \cite{Follana:2007uv},
MILC'10 \cite{Bazavov:2010hj},
ALV'08 \cite{Aubin:2008ie},
RBC-UKQCD'10 \cite{Aoki:2010dy},
PACS-CS'09 \cite{Aoki:2008sm},
PACS-CS'10 \cite{Aoki:2009ix},
BMWc'10 \cite{Durr:2010hr},
QCDSF-UKQCD'10 \cite{Horsley:2010xx} and
JLQCD/TWQCD'09 \cite{Noaki:2009sk}.
}
\end{figure}

Fig.~\ref{fig:fkfpi} summarizes recent lattice determinations of the
decay constant ratio $f_K/f_\pi$. There is a remarkable overall
consistency both among the different determinations and between the
direct lattice determinations and the standard model expectation.

\section{Concluding remarks}

It is fair to say that in recent years lattice QCD has fulfilled one of
its original promises - the post-diction of the ground state light
hadron spectrum. There is a remarkable convergence of results from
different groups using a wide range of lattice actions and analysis
techniques which demonstrates that a large part of the experimentally
observed ground state hadron spectrum can be reproduced within a few
percent accuracy. The most important message from the lattice to the
general particle physics community is that our techniques are in place
and our sources of systematic errors are understood and in some relevant
cases under control. Both central values and total error estimates of
phenomenologically relevant lattice predictions such as those of the
pseudoscalar decay constants can be trusted and contain valuable physics
information. In these areas lattice calculations are preparing to enter
the sub-percent level precision regime with inclusion of subleading
effects such as QED corrections.

In contrast, the current status of excited state hadron spectroscopy is
a reminder that precision calculations are currently possible only for a
small subset of observables that are relatively easy to
compute. Although progress has been impressive, it still
remains a challenging task to obtain a full quantitative understanding
of the excited hadron spectrum.

\section*{Acknowledgements}

I would like to thank C.~Bernard, P.~Boyle, J.~Carbonell, G.~Herdoiza,
S.~D\"urr, Z.~Fodor, R.~Horsley, K.~Jansen, F.~Knechtli, S.~Krieg,
T.~Kurth, C.~Lang, D.~Leinweber, L.~Lellouch, R.~Mawhinney, C.~McNeile,
A.~Ramos, G.~Schierholz, C.~Urbach, S.~Wallace and A.~Walker-Loud for
their support in preparing this review and all my colleagues from the
Budapest-Marseille-Wuppertal collaboration and the transregional
research center ``Hadron physics from lattice QCD'' for innumerable
helpful discussions. Support from the DFG under the grant TR-SFB 55 is
greatly appreciated.

\bibliographystyle{JHEP-3}
\bibliography{references}{}

\end{document}